# DADU: Dual Attention-based Deep Supervised UNet for Automated Semantic Segmentation of Cardiac Images

Racheal Mukisa and Arvind K. Bansal

Department of Computer Science
Kent State University, Kent, OH 44242, USA
{rmukisa1, akbansal}@kent.edu

**Abstract.** We propose an enhanced deep learning-based model for image segmentation of the left and right ventricles and myocardium scar tissue from cardiac magnetic resonance (CMR) images. The proposed technique integrates UNet, channel and spatial attention, edge-detection based skip-connection and deep supervised learning to improve the accuracy of the CMR image-segmentation. Images are processed using multiple channels to generate multiple feature-maps. We built a dual attention-based model to integrate channel and spatial attention. The use of extracted edges in skip connection improves the reconstructed images from feature-maps. The use of deep supervision reduces vanishing gradient problems inherent in classification based on deep neural networks. The algorithms for dual attention-based model, corresponding implementation and performance results are described. The performance results show that this approach has attained high accuracy: 98% Dice Similarity Score (DSC) and significantly lower Hausdorff Distance (HD). The performance results outperform other leading techniques both in DSC and HD.

## 1 Introduction

Cardiovascular diseases are one of the major causes of mortality and morbidity globally, according to the World Health Organization, resulting in more than 18 million deaths every year worldwide [1]. In addition, human suffering, especially among the aging population, requires significant human and other resources, including excessive cost of cardiac surgery to care for the patients. This suffering and lack of resources can be significantly reduced using early detection of cardiac diseases.

In addition to irregular heartbeats (arrhythmia), major fraction of cardiac diseases are classified as defects in blood-channels or perforations in septum causing blood flow problems or mix-up of oxygenated and deoxygenated blood in left and right chambers; defects in cardiac-valve functionality or defects in papillary muscles involved in valvular leaflet motion, especially in older adults, causing asynchrony in valvular motion resulting in blood-regurgitation; blood-flow blockage due to deposits in blood-channels causing lower blood flow, hypertension, myocardial hypertrophy, and/or ischemia that ultimately results in myocardial infarction and mortality [2, 3].

Medical imaging techniques play a significant role in the diagnosis and management of cardiac abnormalities. Cardiac imaging techniques such as Echocardiography (EchoCG), Cardiac Magnetic Resonance (CMR) and Computed Tomography (CT) are extensively used to localize and diagnose cardiac defects [3, 4]. CMR has higher resolution and reproducibility than other imaging modalities [3–5].

Accurate delineation of the left ventricle (LV), right ventricle (RV), and myocardial wall (LMyo) for end diastolic (ED) and end systolic (ES) phases is a crucial step to analyze the cardiac function and defects. Manual delineation of cardiac structures using short-axis cine images has been the clinical practice for cardiac disease diagnosis. This method is prone to errors due to inter-observer variability in experts' assessment and human errors, including fatigue-related errors [6, 7].

The development of efficient and automated methods for cardiac segmentation of CMR and the careful delineation of cardiac structures, such as LV, RV and the myocardium scars, is the first critical step for accurate computation of cardiac parameters, such as the ejection fraction of blood, LV and RV volumes at both end diastole and end systole [5]. Automated cardiac image segmentation involves breaking down the image into semantically meaningful regions, based on which quantitative measures such as LV volume, RV volume, myocardial mass, wall thickness and ejection fraction (EF) can be extracted and analyzed [5, 6].

Lately, deep learning (DL) based semantic segmentation techniques are being explored to assist medical experts diagnose and make informed medical decisions regarding interventions and treatments. Convolution-based techniques aggregate local features; Deeply supervised models reduce information-loss between the comprising hidden layers in the deep learning network; Attention-based models (including transformers) exploit relationships between comprising image-patches to improve accuracy [7-19].

Deep learning techniques, such as Convolution Neural Net (CNN) based techniques like UNet, deeply supervised models, and attention-based techniques, including channel and spatial attention have recently been used for automated semantic segmentation of medical images. CNN based techniques aggregate local features; Deeply supervised models reduce information-loss between the comprising hidden layers in the deep learning network; Attention-based models (including transformers) exploit relationships between comprising image-patches, providing global dependencies between features to improve accuracy [7-19].

Previous research on automated CMR image-segmentation techniques belongs to four major categories: 1) region and edge-based methods; 2) prior shape-based models; 3) deformable model methods, such as the active contour model; 4) deep learning semantic-segmentation models exploiting CNN, UNET, self-attention, and their integrations. While these techniques provide good image-segmentation, they lack versatility of the visual-attention based image-segmentation and analysis used by human experts. For automated systems to become effective, the segmentation process needs to augment deep learning techniques with human-like visual attention.

Human visual attention is a complex integration of 1) top-down task-oriented segmentation for efficient target-localization, focusing, and pruning the unwanted image-space; 2) bottom-up approach to integrate local and global features' analysis, based on spatial proximity, similarity, and relationship between image-patches; 3) adoptive temporal fixation of gaze for adoptive feature-selection and patch-size adjustment for finer analysis of local features; 4) concurrent processing of selected features and fusing the derived information for efficiency [20-22].

Recent developments in self-attention-based vision transformer (ViT) and its variants have partially simulated the effect of bottom-up human visual attention [23]. The mechanism splits an image into a sequence of smaller-size patches and derives similarity-based relationship between these patches to generate embeddings, encoded as a matrix. This matrix is matched for similarity from a large database of pre-trained embeddings derived from millions of images.

In recent years, there have been approaches to integrate convolution-based local features with self-attention-based relationship between image-patches to improve image classification [16, 17]. Integrating convolution with channel and spatial attention-based models, such as CBAM [24], along with reducing computational overhead techniques such as regularization and dropouts can significantly improve accuracy of image segmentation [24-28]. The use of channel and spatial attention is a step closer to visual attention used by human experts.

CBAM based channel and spatial attention and research on deep supervision to reduce information loss and improve vanishing gradient problem has influenced this research [24, 29]. We integrate and extend U-Net with edge-based skip-connection, CBAM-based channel and spatial attention, and deep supervision to improve the accuracy of automated CMR image-segmentation, while reducing the resource requirements.

This research proposes a <u>D</u>ual <u>A</u>ttention-based <u>D</u>eep supervised <u>U</u>Net (DADU) model for an automated segmentation of CMR-images to identify LV, RV, and LMyo. DADU integrates DenseNet, channel and spatial attention with convolution, deep supervision, and actual contour extraction from encoders for skip-connections. It improves feature-selection by applying the channel and spatial attention mechanism coupled with contour detection-based skip connections and a deep supervision strategy for improved accuracy and reduced computational overhead.

The major contributions of this research are as follows:

(1) Integration of channel attention, spatial attention, and real edge extraction as skip-connections with U-Net for accurate CMR image-segmentation;
(2) Building deeply supervised auxiliary paths to reuse network-features and reduce vanishing gradient problem.

The paper is organized as follows. Section 2 describes notation conventions; and background concepts. Section 3 discusses related work. Section 4 describes our approach. Section 5 describes the implementation. Section 6 discusses the results. Section 7 discusses limitations and future work.

## 2   Background

### 2.1   Notations

We use the following mathematical notations in this paper. The symbol $\mathbb{R}$ denotes a real-number domain. The symbol $\sigma$ denotes the sigmoid function. The operator $\otimes$ denotes the element-wise multiplication of two matrices or vectors. The operator $\oplus$ denotes the addition of two feature-maps. The symbol $\bullet$ denotes the concatenation of two feature-maps. The symbol '$\rightarrow$' denotes a finite mapping between two sets. The symbol '$\times$' denotes the Cartesian product of two sets. The letter $E$ in a super-scripted feature-variable denotes association with an encoder. The letter $D$ in a super-scripted feature-variable denotes association with a decoder. The presence of the letters '$Ch$' in a super-scripted feature-variable denotes an output after the Channel Attention Module (CAM). The presence of the letters '$Sp$' in a super-scripted feature-variable denotes an output after the Spatial Attention Module (SAM). The super-scripted capital letter $F$ denotes a feature-set output from an encoder or a decoder. The super-scripted capital letter $M$ denotes a merged feature-map. The capital letter $C$ denotes the set of channels-indices. The small letter 'c' denotes a channel-index in the set of channels $C$. $|C|$ shows the size of the set of the channels or the last-index in the set C. The capital letters $H$ and $W$ denotes variables *Height* and *Width* for a feature-map, respectively.

### 2.2   Convolution Network and U-Net

Convolution network comprises a stack of three layers: convolution filters (CF) to extract feature-maps; Rectilinear Error Linear Unit (RELU) or Gaussian Error Linear Unit (GELU) for noise removal; Pooling layer for summarization, followed by a classification layer such as Feed-forward Neural Network (FNN) or SoftMax function. *Fully Convolution Network* (FCN) is a combination of convolution network for feature-map extraction followed by a SoftMax function layer for classification. Convolution Neural Network (CNN) is a combination of convolution network for feature-map extraction followed by an FNN for classification.

UNet is a popular variant of FCN for medical image analysis [15-19]. UNet delivers high segmentation accuracy with a smaller set of training samples, and its decoder transforms high-level features into semantic labels through up-sampling. UNet has a symmetric architecture and comprises a stack of down-sampling encoder-layers for progressive feature-maps extraction followed by up-sampling decoder-layers for reconstructing the image from the feature-maps; skip connections between an encoder layer and the corresponding decoder layer replenish the information lost during down-sampling. Since its inception, the UNet has evolved into UNet++ [19] and attention UNet [17]. UNet++ fuses connections from low-level feature-maps in the encoder and high-level feature-maps from the decoders [19]. Attention UNet combines local-level feature-maps of UNet with dependency present in self-attention to provide better context [17].

### 2.3 Attention Mechanism

Human visual attention focuses and labels a specified object based upon prior knowledge, history, spatial and temporal context, relationship between the objects (or image-patches) in proximity and uses persistence on focused objects for improved finer resolution analysis [20-22]. Visual attention uses contexts for filling in the missing information [20]. Visual attention selectively blocks unwanted signals, concurrently processes multiple selected relevant features, and fuses the derived information for object detection [21]. It uses both top-down and bottom-up strategies [20, 21]. The top-down strategy is task-oriented and prunes the image space. Bottom-up approach analyzes the local and global features and derives the relationship between image-patches along with adoptive feature-selection and image-patch size for finer analysis [20, 21].

Computational attention-based networks predict the most probable missing entity in a sequence of entities using left-sequential or bidirectional context provided by the attention mechanism. Self-attention for vision transformers captures long-range dependencies within the sequence of image-patches of the same image [23]. The attention mechanism improves the accuracy in capturing non-local interactions and context while helping in suppression of irrelevant information [23]. Computational attention partially simulates bottom-up human visual attention. However, it lacks a top-down approach, adoptive feature-selection based finer analysis of image-patches, and temporal persistence in image analysis.

Channel attention focuses the network to understand 'what' feature-map is meaningful for an input image [24]. Channel attention distinguishes the feature-expression capabilities of different channels in the feature-map to emphasize their importance [24, 28]. Emphasis is on meaningful representation of features by assigning a weight to each channel in the feature map: higher weight means higher relevancy.

Spatial attention focuses an algorithm on the 'where' part, such as where is a feature of interest, by enhancing informative features while suppressing the less important ones [26, 28].

### 2.4 Deep Supervision

To boost gradient flow in the network, multiple auxiliary paths are created between intermediate layers and the final output layer. The losses from auxiliary paths are added to derive cumulative loss. The cumulative loss is minimized for faster convergence and better accuracy [29].

## 3 Related Work

Multiple deep learning techniques such as CNN, FCN, U-Net, auto-encoders, vision transformers, their variants, and combinations are being experimented for accurate cardiac segmentation [7-19].

Avendi et al. proposed an integration of CNN and auto-encoders for automated segmentation of the right ventricle (RV) in cardiac short-axis MRI [7]. Their model integrated with deformable models converges faster compared to conventional deformable models. However, the technique does not capture the contours of LV, RV, LMyo accurately as they do not extract edge-attributes. RV-segmentation suffers from irregular shape and surroundings.

Khened et al. have exploited FCN to optimize skip-connections during up-sampling by extracting hand-crafted features from cardiac structures to identify the region of interest [11]. In addition, an ensemble classifier was used to predict cardiac diseases. FCN goes through uncompensated information-loss during down-sampling, which results in the loss of finer details in substructure contours.

Painchaud et al. used a combination of CNN based shape recognition and autoencoder-based constrained shape-morphing of a prior shape contour to derive the segmentation [12]. Zotti et al. combined the knowledge of prior shape of the substructures and CNN to label LV, RV and LMyo [13]. These approaches suffer from approximation inherent in the use of prior shape and shape-morphing.

Grinias et al. exploited Markov random field (MRF) after the estimation of Gaussian probability density functions, and active contours were used between end-diastolic (ED) and end-systolic (ES) phases [14]. This approach suffers from the approximation used in energy minimization-based estimation of contours.

Li et al. proposed a multi-modal multi-tasking method in a variant of UNet to derive LV, RV and LMyo segmentation concurrently using a shared encoder [15]. The derived features are aggregated in the corresponding decoder. The scheme suffers from information loss inherently present in the encoding process. Besides, they do not use attention-based dependency and deep supervision which reduces their accuracy.

Zhao et al. integrated UNet++ with a feature pyramid network to segment coronary arteries in invasive coronary angiography [18]. The integration exploits both local and global features. However, the model suffers from the information-loss and vanishing gradients due to the absence of edge-detection and deep supervision. It also lacks the channel and spatial attention to improve feature selection.

Gun et al. applied attention UNet (AUN) to perform image segmentation of echocardiograms to derive LV, RV, and LMyo segments [19]. AUN adopts self-attention in each UNet layer to suppress irrelevant regions while highlighting salient features in specific local regions. The use of attention gives excellent results, as shown in Section 6 (see Table 3). However, the scheme lacks edge-detection and deep supervision. Hence, the scheme suffers from information-loss.

Our scheme integrates dense blocks to reduce information-loss between encoder layers, uses dual attention blocks (channel and spatial attention) to pick relevant features in specific local regions both in encoder and decoder layers, uses edge-detection for skip connection to reduce information-loss between encoder and decoders, and uses deep supervision to reduce vanishing gradient problem. As shown in Section 6 (see Table 3), this integration provides significantly improved segmentation.

## 4  DADU Architecture

Figure 1 illustrates a schematic of DADU architecture. It comprises a stack of encoder layers augmented with dense blocks (as in DenseNet), a stack of decoder layers, a combination of edge-detection (ED) and Dual Attention Block (DAB) as skip-connections, followed by a deep supervision module comprising multiple auxiliary paths from distinct levels of decoder stage. We replace convolution blocks (as in U-Net) in encoders with dense blocks (as in DenseNet) [21]. Dense blocks use residual links from previous encoder layers along with convolution filters to reduce information-loss. The deep supervision mechanism adds direct supervision to the hidden layers which offers fast convergence and better segmentation accuracy [20]. Deep supervision module also prevents the model from over-fitting during training.

DAB captures information from both encoder and decoder feature-maps to generate attention values. The encoder path extracts context information and encodes input images by repeatedly applying two $3 \times 3$ convolutions and each followed by a batch normalization layer, a rectified linear unit (ReLU) and a $2 \times 2$ max pooling operation. A stride of 2 is used for down-sampling by the pooling layer and the spatial dimensions are decreased gradually while the number of feature channels are doubled at each down-sampled feature map. We apply a $1 \times 1$ convolution operation with sigmoid activation at the end to generate the final segmentation map.

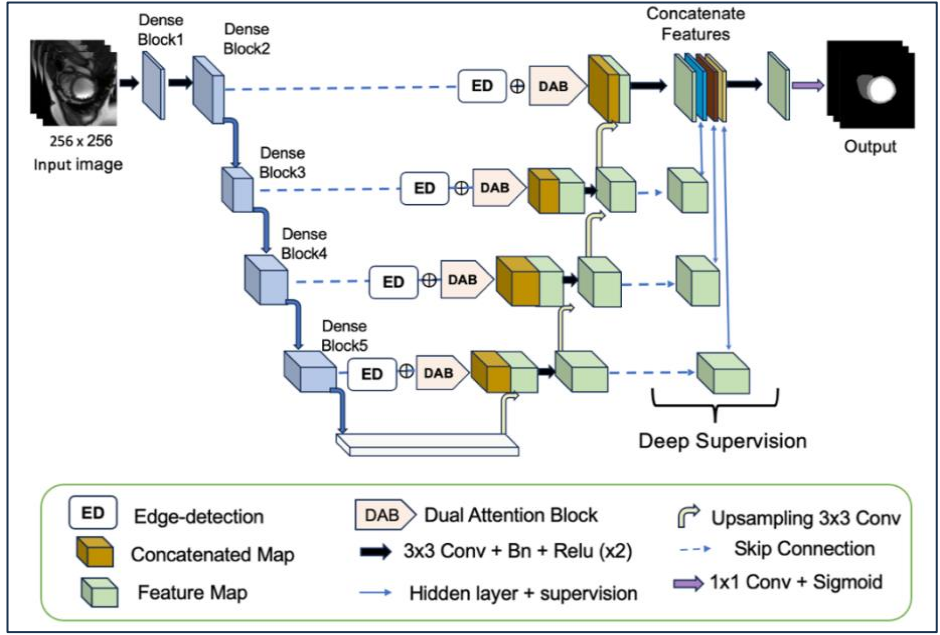

**Fig 1.** A schematics of DADU architecture

Decoder-features are generated by multiplying the encoder features with the Dual Attention Block (DAB), which comprises channel and spatial attention modules, integrated with derived edge-information. Skip-connections include the down-sampled feature-maps from the encoder and the decoder's feature-maps using DAB analysis and edge information. These integrated skip-connections allow corresponding decoders to combine high-level context and low-level feature-maps.

The use of dense blocks, skip-connections and deep supervision module significantly reduces information-loss and creates an effective information-flow through the network by reusing features and enhancing gradient-flow during training. Additionally, deep supervision forces the attention-map of the intermediate features to be discriminative at every decoder-level, improving the overall performance.

### 4.1 Dual Attention Block (DAB)

DAB (see Fig. 2) includes two types of modules: Channel Attention Module (CAM) and the Spatial Attention Module (SAM). CAM enhances the contextual information into the low-level encoder-features to reduce the semantic-gap between the encoder and decoder features. SAM helps the model concentrate on the important region in the image structure. DAB maximizes the benefit of both modules to construct the final refined feature-map $F^{out}$.

DAB seamlessly integrates the strengths of both the channel and spatial attention modules to enhance our model's overall segmentation performance. CAM is applied followed by SAM, based upon an experimental study that this order is computationally more efficient [28]. DAB is added at every feature-map dimension through skip connections to help the network focus on components and more valuable features for the final output.

Given an input feature-map pair, $F^{in} = (F^{inE}, F^{inD})$, such that $F^{inE} \in \mathbb{R}^{C \times H \times W}$ from an encoder, and $F^{inD} \in \mathbb{R}^{C \times H \times W}$ is from the corresponding decoder, the corresponding DAB applies CAM followed by SAM to derive the feature-map $F^{out}$.

CAM constructs a merged channel-attention map $M^{Ch} \in \mathbb{R}^{C \times 1 \times 1}$, which is combined with the residual link $F^{in}$ using element-wise matrix multiplication to derive feature-map $F^{Ch}$. SAM applies spatial attention on $F^{Ch}$ to generate a merged feature-map $M^{Sp}$. The overall attention process is summarized in eqn. (1) - (2). An algorithm for computing DAB is given in Algorithm 1.

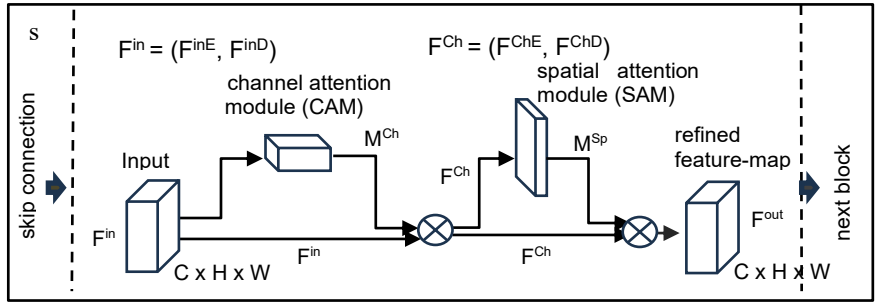

**Fig 2**: An illustration of Dual Attention Block (DAB)

$$F^{Ch} = M^{Ch} \otimes F^{in} \qquad (1)$$

$$F^{out} = M^{Sp} \otimes F^{Ch} \qquad (2)$$

---

**Algorithm 1.** An algorithm to compute final feature-map $F^{out}$ using DAB

---

**Algorithm** DAB

**Input**: 1. feature-map $F^{inE} \in \mathbb{R}^{C \times H \times W}$ of the form $<F_1^{inE}, \ldots, F_C^{inE}>$

2. feature map $F^{inD} \in \mathbb{R}^{C \times H \times W}$ of the form $<F_1^{inD}, \ldots, F_C^{inD}>$

**Output**: Attention-modulated refined feature map $F^{out} \in \mathbb{R}^{C \times H \times W}$

```
{   F^out = { }; % # initialize refined feature map F^out
    ∀i(1 ≤ i ≤ |C|) % process all the channels
    {   get encoder-decoder pair (F_i^inE, F_i^inD) ∈ (F^inE, F^inD);
        F^in = (F_i^inE, F_i^inD);
        {   # construct the refined feature-map using channel and spatial attention
            M^Ch = channel_attention(F^inE, F^inD); % use CAM to build merged feature map M^Ch
            F^Ch = σ (M^Ch ⊗ F^in); % F^Ch is final merged channel attention map
            M^Sp = spatial_attention(F^Ch); % apply SAM to build  merged attention map M^Sp
            F^out = σ(M^Sp ⊗ F^Ch); % build final attention-map by combining F^Ch
        }
    } return (F^out) %  return final-feature map
}
```

---

### 4.2 Channel Attention Module (CAM)

CAM uses max-pooling and average-pooling to highlight informative regions in each channel. *Average-pooling* enables feedback on all points from each feature-map during the gradient backpropagation; *Max-pooling* provides feedback where the response is highest in the feature-map during backpropagation [19]. CAM combines channel information from feature-maps of both an encoder and the corresponding decoder. The decoder provides rich semantic information to capture semantic dependencies.

The input to CAM is a feature-map pair $F^{in} = (F^{inE} \in \mathbb{R}^{C \times H \times W}, F^{inD} \in \mathbb{R}^{C \times H \times W})$. Channel attention is computed by first aggregating spatial information of each encoder and decoder feature-map separately using average-pooling and max-pooling, generating vectors: $F^{inE-avg} \in \mathbb{R}^{C \times 1 \times 1}$ and $F^{inE-max} \in \mathbb{R}^{C \times 1 \times 1}$, respectively. The vectors $F^{inE-avg}$ and $F^{inE-max}$ are added to derive merged feature-map $M_{c \in C}^{ChE}$ (see eqn. 3). Similarly, average-pooling and max-pooling operations work on decoder feature-map $F^{inD}$ to generate two different vectors: $F^{inD-avg} \in \mathbb{R}^{C \times 1 \times 1}$ and $F^{inD-max} \in \mathbb{R}^{C \times 1 \times 1}$. The vectors $F^{inD-avg}$ and $F^{inD-max}$ are added to derive merged feature-map $M_{c \in C}^{ChD}$ (see eqn. 4). The merged attention-map of the encoder features $M_{c \in C}^{ChE}$ and the merged attention-map of the corresponding decoder features $M_{c \in C}^{ChD}$ are added element-wise to derive merged channel attention map $M_{c \in C}^{Ch}$ (see eqn. 5).

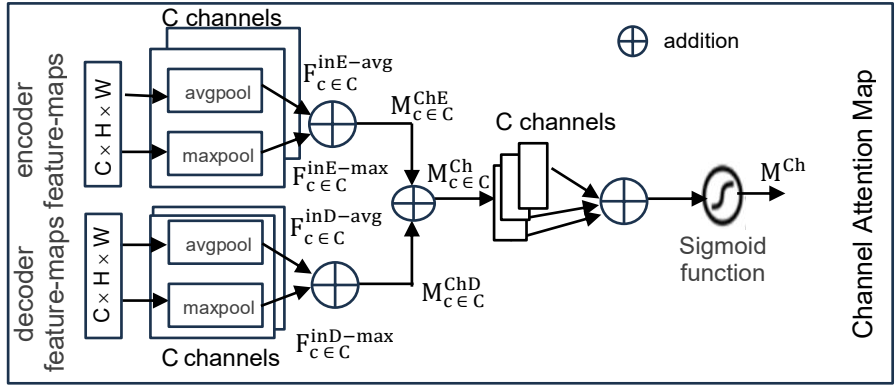

**Fig 3**. A schematics of Channel Attention Module

A sigmoid function is applied on the weighted sum $M^{Ch} = \sum w_c M^{Ch}_{c \in C}$ from all channels in the set of channels $C$ to derive the final vector $M^{Ch}$, which is passed on to the corresponding SAM (see eqn. 6). Algorithm 2 describes an algorithm to derive $M^{Ch}$.

$$M^{ChE}_{c \in C} = F^{inE-avg}_{c \in C} \oplus F^{inE-max}_{c \in C} \qquad (3)$$

$$M^{ChD}_{c \in C} = F^{inD-avg}_{c \in C} \oplus F^{inD-max}_{c \in C} \qquad (4)$$

$$M^{Ch}_{c \in C} = M^{ChE}_{c \in C} \oplus M^{ChD}_{c \in C} \qquad (5)$$

$$M^{Ch} = \sigma(w_1 M^{Ch}_1 \oplus w_2 M^{Ch}_2 \oplus \ldots \oplus w_{|C|} M^{Ch}_{|C|}) \qquad (6)$$

### 4.3 Spatial Attention Module (SAM)

SAM is constructed based on the interrelationship between the spatial information that focuses on the salient regions. Spatial features from different spatial locations are aggregated by adding the dilated convolutions (stride-size > 1). Higher stride-size facilitates capturing non-local features.

As illustrated in Fig. 4, to compute the spatial attention in each encoder and decoder feature, average-pooling operation, max-pooling operation, and 1×1 convolution operation are applied along each channel axis and the outputs are concatenated to generate vectors: $F^{Sp-avg} \in \mathbb{R}^{1 \times H \times W}$, $F^{Sp-max} \in \mathbb{R}^{1 \times H \times W}$ and $F^{Sp-1\times1} \in \mathbb{R}^{1 \times H \times W}$, which denote the map generated from average-pooled features, max-pooled features and 1×1 convolution, respectively. The encoder feature contains rich location information and helps the model to focus on salient regions beneficial for finding the object's location and determining the target-structure in the image. $F^{Sp-avg}$, $F^{Sp-max}$, $F^{Sp-1\times1}$ are concatenated, and a convolution layer with kernel size 7×7 is applied to generate a concatenated spatial attention-map $M^{SpE}_{c \in C} \in \mathbb{R}^{1 \times H \times W}$ for an encoder and $M^{SpD}_{c \in C} \in \mathbb{R}^{1 \times H \times W}$ for the corresponding decoder in each channel.

**Algorithm 2.** An algorithm for Channel Attention Module
---

**Algorithm** channel_attention
**Input**: 1. feature-map $F^{inE} \in \mathbb{R}^{C \times H \times W}$ of the form $<F_1^{inE}, \ldots, F_{|C|}^{inE}>$
        2. feature map $F^{inD} \in \mathbb{R}^{C \times H \times W}$ of the form $<F_1^{inD}, \ldots, F_{|C|}^{inD}>$
**Output**: Channel attention map $M^C \in \mathbb{R}^{C \times 1 \times 1}$

{ $M^{Ch}$ = {}; % initialize the attention map
  { $\forall i\ (1 \leq i \leq |C|)$
      { get encoder-decoder pair $(F_i^{inE}, F_i^{inD}) \in (F^{inE}, F^{inD})$;
      # Apply average and max pooling and merge
      $F_i^{inE-avg}$ = avgPool($F_i^{inE}$); $F_c^{inE-max}$= maxPool($F_i^{inE}$);
      $F_i^{inD-avg}$ = avgPool($F_i^{inD}$); $F_c^{inD-max}$= maxPool($F_i^{D}$);
      $M_i^{ChE}$ = $F_i^{inE-avg} \oplus F_i^{inE-max}$;    $M_i^{ChD}$ = $F_i^{inD-avg} \oplus F_i^{D-max}$;
      $M_i^{Ch} = M_i^{ChE} \oplus M_i^{ChD}$; % add encoder merged map with decoder merged map
      $M^{ch} = M^{ch} \oplus M_i^{Ch}$; % iteratively build merged map $M^{ch}$
    } $M^{ch} = \sigma(M^{ch})$; % apply sigmoid function
  } **return** ($M^{Ch}$);
}

These concatenated spatial attention maps from encoder and decoder are added to derive merged feature-map $M_{c \in C}^{Sp}$ for each channel c ∈ C. The merged feature-maps from all the channels are added, and a sigmoid activation function is applied on the summation to calculate the merged spatial attention-map $M^{Sp}$. Equations (7)-(10) describe spatial attention. The function $f^{7 \times 7}$ represents a convolution operation with the kernel-size of 7 × 7. An algorithm for SAM module is given in Algorithm 3.

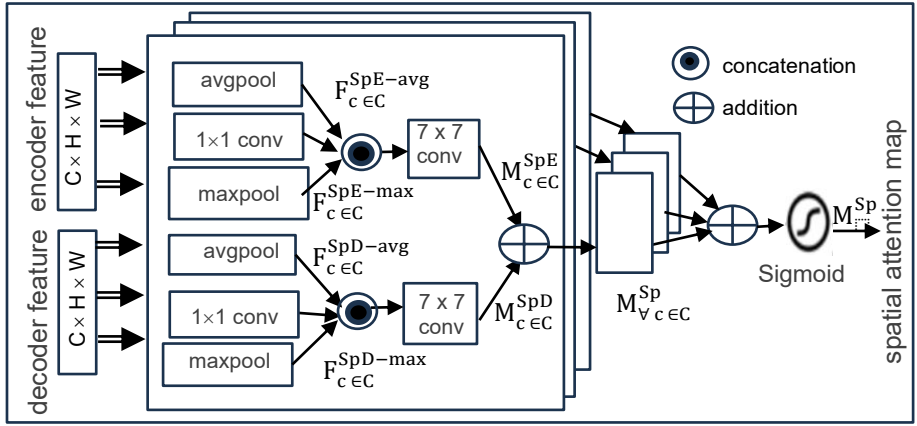

**Fig 4.** A schematics of spatial attention module

$$M_{c \in C}^{SpE} = f_{x,y}^{7 \times 7}\left(\left[F_{c \in C}^{ChE-avg} \bullet F_{c \in C}^{ChE-1\times 1} \bullet F_{c \in C}^{ChE-max}\right]\right) \quad (7)$$

$$M_{c \in C}^{SpE} = f_{x,y}^{7 \times 7}\left(\left[F_{c \in C}^{ChE-avg} \bullet F_{c \in C}^{ChE-1\times 1} \bullet F_{c \in C}^{ChE-max}\right]\right) \quad (8)$$

$$M_{c \in C}^{SpE} = M_{c \in C}^{SpE} \oplus M_{c \in C}^{SpE} \quad (9)$$

$$M^{Sp} = \sigma(M_1^{Sp} \oplus M_2^{Sp} \oplus \ldots \oplus M_{|C|}^{Sp}) \quad (10)$$

---

**Algorithm 3.** An algorithm for SAM

---

**Algorithm** spatial_attention
**Input**: 1. feature-map $F^{ChE} \in \mathbb{R}^{1 \times H \times W}$ of the form $<F_1^{ChE}, \ldots, F_{|C|}^{ChE}>$

2. feature map $F^{ChD} \in \mathbb{R}^{1 \times H \times W}$ of the form $F_1^{ChD}, \ldots, F_{|C|}^{ChD}>$

**Output**: Spatial attention map $M^{Sp} \in \mathbb{R}^{C \times H \times W}$
{ $M^{Sp}$ = {}; % initialize the attention map
  $\forall i \ (1 \leq i \leq |C|)$
  { get encoder-decoder pair $(F_i^{ChE}, F_i^{ChD}) \in (F^{ChE}, F^{ChD})$;
    $\forall$ spatial location (x, y)
    { # Apply average, max pooling and 1x1conv and concatenate.
      $F_i^{SpE-avg}$ = avgPool($F_i^{ChE}$); $F_i^{SpD-max}$ = maxPool($F_i^{ChE}$);
      $F_i^{SpE-1x1}$ = $f^{1\times 1}(F_i^{ChE})$;
      $F_i^{SpD-avg}$ = avgPool($F_i^{ChD}$); $F_i^{SpD-max}$ = maxPool($F_i^{ChD}$);
      $F_i^{SpD-1x1}$ = $f^{1\times 1}(F_i^{ChD})$;
    }
    $M_i^{SpE} = F_i^{SpE-avg} \bullet F_i^{SpE-max} \bullet F_i^{SpE-1x1}$; % concatenate
    $M_i^{SpD} = F_i^{SpD-avg} \bullet F_i^{SpD-max} \bullet F_i^{SpD-1x1}$; % concatenate
    $M_i^{Sp} = M_i^{SpE} \oplus M_i^{SpD}$;  $M^{Sp} = M^{Sp} \oplus M_i^{Sp}$;
  }
  $M^{Sp} = \sigma(M^{Sp})$; **return**($M^{Sp}$);
}

### 4.4 Deep Supervision Mechanism

To accurately segment cardiac structures in CMR images from an already complex heart anatomy, deep models are needed to encode highly representative features. Enhancing the gradient flow for shallow layers with the deep supervision mechanism has proven to be effective in improving the segmentation performance [29]. Output is generated at each level. The loss is propagated backwards to the deeper layers during parameter updates to improve the overall performance. At each decoding branch level, the input feature-map is processed using convolution and interpolation for up-sampling. The sigmoid function is used to calculate the probability map of segmentation results. The losses of the auxiliary paths from contributing layers and the final output layer are combined to compute the final loss $\mathcal{L}$ as shown in Fig. 5.

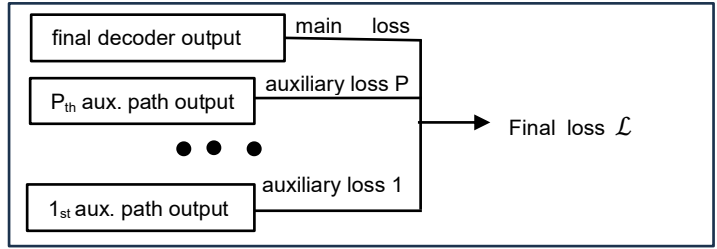

**Fig 5**. A schematics of the deep supervision links to compute the total loss.

In eqn. 11, the symbol $\mathcal{X}$ denotes the input vector. The symbol $W = \{w_1, w_2, \ldots, w_P\}$ denotes the vector of weights for auxiliary paths, where denotes the weight of the $i^{th}$ auxiliary path. The symbol $P$ denotes the number of auxiliary paths. The symbol $\eta_s$ denotes the balancing weight of $\mathcal{L}_s$, which is adjusted during backpropagation. Weights are adjusted during backpropagation to minimize the final loss $\mathcal{L}$. The term $\tilde{\mathcal{L}}(\mathcal{X}, W)$ denotes the contributed loss from the final decoding layer.

The final loss $\mathcal{L}(\mathcal{X}, W, \eta_s)$ is calculated using eqn. 11. Final loss is the weighted sum of the 1) $\tilde{\mathcal{L}}(\mathcal{X}, W)$ − loss calculated from the final decoder; 2) $\sum_{i \leq P} \eta_s \mathcal{L}_s(\mathcal{X}, w_i \in W)$ − cumulative loss from all auxiliary paths.

The final loss is used for backpropagation to all the contributing layers to reduce the influence of vanishing gradient and improve the model's learning efficiency and optimization.

$$\mathcal{L}(\mathcal{X}, W, \eta_s) = \tilde{\mathcal{L}}(\mathcal{X}, W) + \sum_{i \leq P} \eta_s \mathcal{L}_s(\mathcal{X}, w_i \in W) \qquad (11)$$

## 5  Implementation

The software was written in Python using the PyTorch framework. The program was executed on a Dell server with a GPU GEForce RTX 2070 system. The model was trained using input images alongside their corresponding masks. We employed a fivefold cross-validation approach. The model was optimized via the ADAM optimizer with a learning rate of 0.001. We computed the loss using minimum batch gradient and a batch size of 10. Training concluded after 100 epochs, with the best-performing models preserved for subsequent testing. Our attention model takes 2.2 hours to train using 4GB of GPU memory. During testing, it takes 0.2 seconds and 1.5 GB GPU memory to generate a single output for the entire phase, which shows that our model is light-weight, efficient to train and has potential for easy deployment in a clinical setting.

### 5.1  Dataset

The implementation was evaluated using the ACDC dataset that included LV, RV, and LMyo information of 150 examinations from different cardiac patients [30, 31]. The data comprised short-axis cine-MRI taken on both 1.5T and 3T systems with resolutions ranging between 0.70 mm × 0.70 mm to 1.92 mm × 1.92 mm in-plane and a range through-paneling from five millimeters (mm) to ten mm thickness. The dataset comprised equally distributed healthy condition (NOR) and four pathological conditions: *myocardial infarction* (MINF), *hypertrophic cardiomyopathy* (HCM), *dilated cardiomyopathy* (DCM), *abnormal right ventricle* (ARV).

We divided the dataset into a training set comprising 100 examinations and a testing set comprising 50 examinations, with manual annotations of the LV, RV, and LMyo provided for both end-systole (ES) and end-diastolic (ED) phases by clinical experts.

### 5.2 Pre-processing

The ACDC dataset contained patients' data in 3D NIFTI21 format. The images were transformed into PNG format. The original data contained severe slice misalignments and varying resolutions due to different breath-hold positions during image-capture. Data was pre-processed to correct these inconsistencies. 3D images were converted to 2D format by taking all the slices for both the original image and the mask as input to ensure that all gray-scale images have a similar voxel-size. The resulting image-slices were pre-processed to derive the corresponding 2D images in gray-scale, scaled to 256 pixels x 256 pixels.

### 5.3 Performance Metrics

The segmentation accuracy was evaluated using the mean *Dice Similarity Coefficient* (*DSC*) and *Hausdorff Distance* (*HD*). *DSC(X, Y)* measures the percentage overlap between the ground-truth mask *X* and the segmentation output *Y* (see eqn. 12). We used *DSC* to compute the loss during the experiments. Higher DSC is better as it shows significant overlap between the original mask and the predicted segment.

$$DSC = \frac{2 \times |X \cap Y|}{|X| + |Y|} \tag{12}$$

*HD(X, Y)* measures the maximum of the minimum of distances from a set of points $y \in Y$ in one contour to set of ground-truth contour-points $x \in X$ (see eqn. 13). A smaller *HD* shows that segmented contour is more aligned with the ground-truth contour.

$$HD(X,Y) = \max_{x \in X} \min_{y \in Y} \| x - y \| \tag{13}$$

## 6 Results and Discussion

Figure 6 shows segmentation predictions for combined segmentation of LV, RV and LMyo. Table 1 shows the average for DSC and HD for the entire dataset. Table 2 shows the segmentation accuracy in End-Diastole(ED) phase and End-Systole(ES) phases for multiple epochs.

Results from Tables 1 and 2 show that segmentation is better in ED phase than in ES phase. The mean HD values are low, signaling better alignment between the ground-truth and the predicted output.

**Table 1.** Segmentation output at End-diastole (ED) and End-systole (ES) for entire dataset.

|              | DSC  |      |      | HD   |      |      |
|--------------|------|------|------|------|------|------|
|              | LV   | RV   | LMyo | LV   | RV   | LMyo |
| **End-diastole** | 0.98 | 0.97 | 0.93 | 1.9  | 10.0 | 8.3  |
| **End-systole**  | 0.96 | 0.94 | 0.92 | 3.9  | 4.0  | 6.4  |

**Table 2.** Epoch-based segmentation results at End-diastole (ED) and End-systole (ES)

| Eps | ED   |      |      |      |      |      | ES   |      |      |      |      |      |
|-----|------|------|------|------|------|------|------|------|------|------|------|------|
|     | DSC  |      |      | HD   |      |      | DSC  |      |      | HD   |      |      |
|     | LV   | RV   | LMyo | LV   | RV   | LMyo | LV   | RV   | LMyo | LV   | RV   | LMyo |
| 20  | 0.93 | 0.89 | 0.84 | 4.1  | 20.6 | 14.6 | 0.89 | 0.88 | 0.89 | 10.0 | 4.2  | 7.4  |
| 40  | 0.95 | 0.92 | 0.86 | 2.2  | 18.3 | 10.2 | 0.95 | 0.90 | 0.90 | 5.8  | 5.7  | 4.9  |
| 60  | 0.96 | 0.95 | 0.86 | 2.8  | 18.6 | 11.4 | 0.96 | 0.91 | 0.89 | 5.6  | 4.5  | 7.2  |
| 80  | 0.98 | 0.97 | 0.86 | 2.4  | 17.3 | 10.4 | 0.96 | 0.92 | 0.91 | 4.0  | 4.6  | 6.8  |
| 100 | 0.98 | 0.97 | 0.92 | 1.8  | 16.4 | 7.6  | 0.96 | 0.94 | 0.91 | 3.0  | 3.9  | 6.4  |

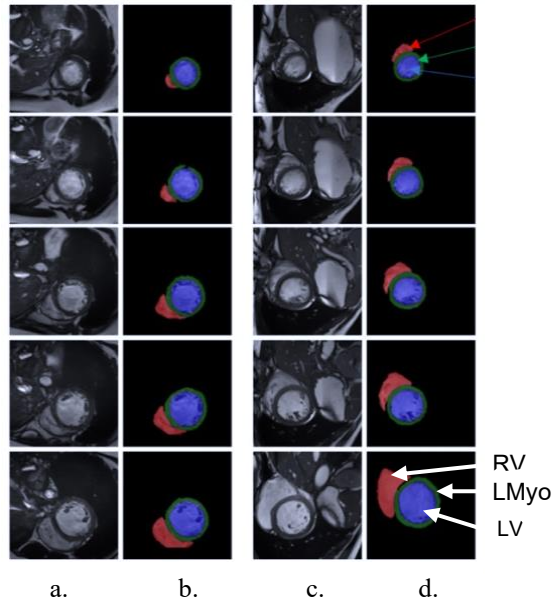

a.　　　b.　　　c.　　　d.

**Fig. 6.** Sample segmentation output from the DADU model. Column a) and c) show the original images, Column b) and d) show the segmented image. All rows from top-to-bottom show the sequential arrangement of images from apex to the base.

Table 2 shows that DSC and HD-values improve with larger epochs. While DSC started saturating around 100 epochs, HD-values progressively improve. However, there is an anomaly around the 40 epochs when HD-values suddenly dip and then, again increase before dipping significantly at 100 epochs.

LV has the best DSC since LV shape in the short-axis CMRs is clearly defined as compared to RV which has crescent shape and thinner wall. HD-values are significantly lower for LV than for RV and LMyo, due to better contrast between the LV and the surrounding LMyo that results in accurate edge-detection. The segmentation of LMyo is more challenging, compared to LV and RV, as shown by lower DSC for both ED and ES.

Figure 7 shows comparison of DADU with other leading segmentation techniques. Visual comparison clearly shows significantly improved segmentation in our approach for all the three structures. UNet performed the worst with an almost invisible LMyo ring. The UNet++ and Attention UNet resulted in partial LMyo ring. Our method formed a complete LMyo, due to the use of dual attention block that focuses on key features, their localization, and deep supervision.

Figure 8 shows comparison of our model's segmentation results with other state-of-the-art methods at ES phase. Rows from top-to-bottom indicate the image slices from apex to basal. Columns from left-to-right show original image, ground-truth mask, our model (DADU), UNet, UNet++ and Attention UNet based segmentation, respectively.

Table 3 shows DSC and HD of ED and ES phases for the comparison of DADU with other leading models. Attention UNet (AUN) and DADU outperform other models. It is our conjecture that integrating attention provides better feature-maps and reduces information-loss. Our model (DADU) outperforms all other models, including AUN, both in DSC and HD. The improvement in HD is significant. There are varying results for RV, but DADU performed closest to the ground truth.

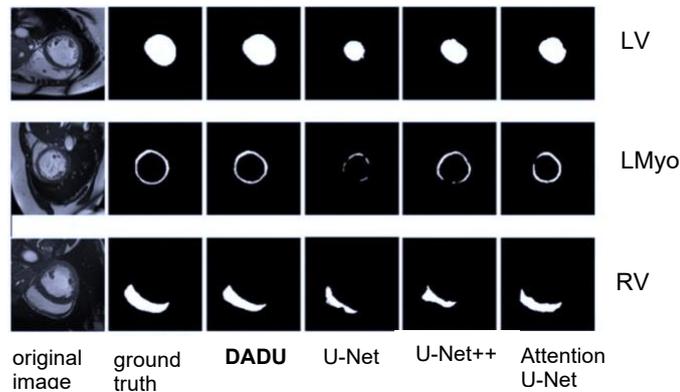

**Fig. 7.** Qualitative comparison of DADU with other state-of-the art methods

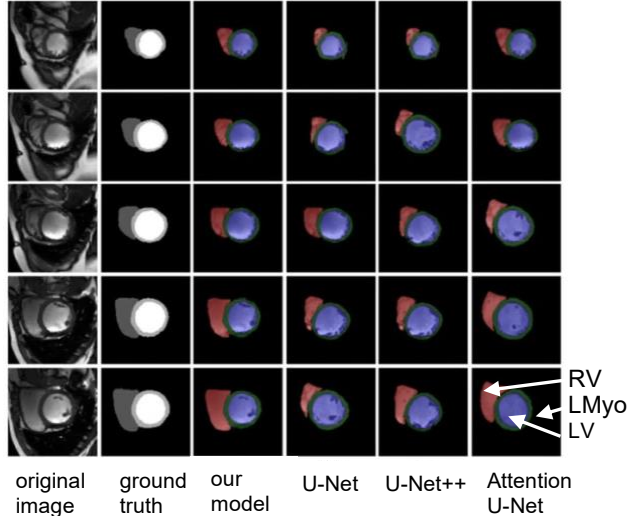

**Fig. 8.** Comparing DADU with state-of-the-art methods in ES phase

Table 3. Comparing DSC and HD values for LV, RV and LMyo with other models

| Methods | LV | | | | RV | | | | LMyo | | | |
|---|---|---|---|---|---|---|---|---|---|---|---|---|
| | DSC | | HD (mm) | | DSC | | HD (mm) | | DSC | | HD (mm) | |
| | ED | ES | ED | ES | ED | ES | ED | ES | ED | ES | ED | ES |
| U-Net [7] | 0.96 | 0.91 | 6.5 | 7.6 | 0.93 | 0.88 | 22.3 | 8.9 | 0.89 | 0.82 | 12.2 | 24.3 |
| FCN [11] | 0.96 | 0.92 | 8.1 | 9.0 | 0.94 | 0.88 | 14.0 | 13.9 | 0.89 | 0.90 | 9.8 | 12.6 |
| CNN [12] | 0.96 | 0.91 | 6.1 | 8.3 | 0.93 | 0.88 | 13.7 | 13.3 | 0.88 | 0.90 | 8.6 | 9.6 |
| CNN [13] | 0.96 | 0.91 | 6.6 | 8.7 | 0.94 | 0.88 | 10.3 | 14.0 | 0.89 | 0.90 | 9.6 | 9.3 |
| MRF [14] | 0.95 | 0.85 | 9.0 | 13.0 | 0.90 | 0.77 | 19.0 | 24.2 | 0.80 | 0.78 | 12.3 | 14.6 |
| UNET++ [18] | 0.96 | 0.94 | 5.5 | 6.4 | 0.94 | 0.92 | 19.3 | 9.3 | 0.90 | 0.90 | 8.5 | 13.8 |
| AUN [19] | 0.97 | 0.94 | 2.9 | 5.6 | 0.95 | 0.93 | 16.6 | 8.8 | 0.91 | 0.91 | 9.0 | 7.0 |
| **DADU** | **0.98** | **0.96** | **1.9** | **3.9** | **0.97** | **0.94** | **10.0** | **4.0** | **0.93** | **0.92** | **8.2** | **6.3** |

## 7. Conclusion, Limitations, and Future Work

Integration of convolution, DenseNet, DAB with extracted edges as skip-connections, and supervised deep learning enhances the segmentation accuracy (especially the Hausdorff distance) of CMR significantly. DAB enhances contextual information within encoder features; SAM ensures that the model focuses on important regions in the short-axis cine CMR, improving segmentation accuracy. Deep supervision strategy reduces the vanishing gradient problem improving the learning rate. Our results show that the proposed model outperforms other leading models.

The model still does not match the intricacies of human visual attention, which combines top-down task specific and bottom-up adoptive finer feature-map analysis. Task-based focus will improve accuracy further and will be needed for abnormality analysis in heart-components [32, 33].

Currently, we are extending our research for automated diagnosis of different cardiac abnormalities and improving our model, taking inspiration from human visual attention [20−22].